\documentclass[12pt]{article}

\usepackage{a4wide,amsfonts,amssymb}
\usepackage{graphics}

\newtheorem{theorem}{Theorem}
\newtheorem{proposition}[theorem]{Proposition}
\newtheorem{lemma}[theorem]{Lemma}
\newtheorem{corollary}[theorem]{Corollary}
\newtheorem{remark}[theorem]{Remark}

\def\Y{\mathbb{Y}}
\def\A{\mathbb{A}}
\def\L{\mathbb{L}}

\def\S{\mathbb{S}}
\def\D{\mathbb{D}}

\def\R{\mathbb{R}}

\def\X{\mathbb{X}}
\def\K{\mathbb{K}}

\begin{document}

\title{A new method to construct spacetimes \\ with a spacelike circle action \thanks{This work was partially supported by MEC-FEDER Grant
MTM2004-04934-C04-01. \newline {2000  {\it Mathematics Subject
Classification.} 53C50.}\newline {\it Keywords.} Lorentzian
manifolds, Null congruence, Semi-Riemannian submersions, isometric
spacelike circle action.}}

\author{Stefan Haesen\footnote{Postdoctoral researcher of the F.W.O.-Vlaanderen} \and Francisco J. Palomo \and Alfonso Romero}
\maketitle

\begin{abstract}
A new general procedure to construct realistic spacetimes is
introduced. It is based on the null congruence on a time-oriented
Lorentzian manifold associated to a certain timelike vector field.
As an application, new examples of stably causal Petrov type D
spacetimes which obey the timelike convergence condition and which
admit an isometric spacelike circle action are obtained.
\end{abstract}


\section{Introduction}

The assumption of the existence of symmetries has been used
extensively to obtain exact solutions of very complicated
equations in Physics. In particular, there are remarkable examples
of exact solutions of the gravitational field equations which have
been obtained by assuming the existence of an isometric action of
a Lie group on the spacetime. On the other hand, if the orbit
space is a (smooth) manifold, many physical and mathematical
problems can be reduced to questions on the orbit space which has
a lower dimension than the original spacetime. We recall for
instance the spatial spherical symmetry for the classical
Schwarzschild solution.

The simplest compact Lie group is the circle $\S^{1}$ and so an
isometric circle action could be the simplest symmetry assumption
on a Lorentzian manifold. Classically, the case of $\S^{1}$ acting
by timelike isometries is well known, recall for example the
timelike circle action admitted for the $3$-dimensional anti-de
Sitter spacetime. Note that a timelike circle action implies the
existence of closed timelike curves and so the absence of the
chronology condition. Recently the case of $\S^{1}$ acting by
spacelike isometries has received a deep and wide attention. In
fact, Choquet-Bruhat and Moncrief \cite{CB,CBM} have introduced a
Lorentzian metric on each $4$-dimensional manifold of the form
$\Sigma \times \mathbb{R} \times \mathbb{S}^1$, where $\Sigma$ is
certain compact orientable $2$-dimensional manifold, which is
invariant under the action of the group $\mathbb{S}^1$ and whose
orbits are spacelike.

This paper is devoted to the introduction of a new general
technique to construct 4-dimensional spacetimes which admit a
spacelike circle action starting from certain 3-dimensional
Lorentzian manifolds. Moreover, we will show that in several
relevant cases this action is isometric. We would like to point
out that every orientable $3$-dimensional manifold can be endowed
with a Lorentzian metric. Further, let us recall that {\it the
space of all Lorentzian metrics on a $3$-dimensional orientable
compact manifold, for a natural topology, possesses an infinity of
connected components} \cite{Mou}. Therefore we think that a method
to construct spacetimes from $3$-dimensional Lorentzian manifolds
would be interesting by itself and it will produce many new
examples of 4-dimensional spacetimes.

The technique used here is based on the geometry of the null
congruence associated to a timelike vector field on a
time-oriented Lorentzian manifolds (see Section 2 for the
definition and main properties of the null congruence). Given an
$n(\geq 3)$-dimensional time-oriented Lorentzian manifold $(M,g)$
and a timelike vector field $K$ on $M$, the null congruence
obtained from this manifold , $C_KM$, is a codimension-two
oriented submanifold of the tangent bundle $TM$. It is constructed
by taking a particular $(n-2)$-dimensional ellipse in the light
cone above every point of $M$ and it can be viewed as the manifold
of all null directions on the Lorentzian manifold $(M,g)$. The
null congruence has been previously used to characterize
Friedmann-Lema\^itre-Robertson-Walker spacetimes \cite{harris}, to
study infinitesimal null isotropy \cite{koch} and recently to
analyze the behavior of conjugate points along null geodesics
\cite{gut1,gut2}, (see also \cite{GPR1,GPR2,PR} for more details).
Moreover, in an appropriate sense, the null congruence allows one
to study the null sectional curvature as a smooth function on the
set of degenerate tangent planes \cite{P}. We present now another
application of a different nature. In fact, every null congruence,
$C_KM$, inherits a natural Lorentzian metric, $\widehat{g}$, from
the well-known Sasaki one on $TM$ defined from $g$, see for
instance \cite{Sak97}, which is indefinite with index $2$. The
canonical projection $\pi : C_KM \rightarrow M$ becomes then a
semi-Riemannian submersion and a fibre bundle with fibre-type a
spacelike $(n-2)$-dimensional sphere \cite{gut1}. In particular,
starting from a time-oriented, $3$-dimensional Lorentzian manifold
$(M,g)$, we obtain a $4$-dimensional spacetime
$(C_KM,\widehat{g})$ which is a fibre bundle with spacelike circle
fibres. It should be noted that $\widehat{g}$ depends on the
metric $g$ and the choice of a timelike vector field $K$ on $M$.
Therefore, this procedure provides a wide family of spacetimes
and, as it will be shown, with a rich variety of geometric
properties.

The content of this paper is organized as follows. Section $2$ is
devoted to introduce basic results and several definitions about
null congruences. Then, the vertical and the horizontal subspaces
for the semi-Riemannian submersion $\pi:C_{K}M\rightarrow M$ are
found, and as a consequence the horizontal lift of a vector field
on $M$ along this submersion is given in (\ref{eqcor1}). It should
be pointed out that the horizontal vectors for the null congruence
are also horizontal for the tangent bundle only if the considered
timelike vector field $K$ is parallel. As an easy consequence, we
can show that the null congruences inherits certain causality
conditions imposed on the base spacetime. In fact, in causality
theory, the stable causality condition is frequently used to
describe suitable spacetimes. As shown in Remark 2, if we assume
that $(M,g)$ is stably causal, then every null congruence over $M$
is also stably causal.

In Section 3, basic relations between the geometries of the
Lorentzian manifold and the null congruence constructed over it
are expressed from the O'Neill fundamental tensors $T$ and $A$
\cite{oneill} of the semi-Riemannian submersion. The formulas
obtained for $T$ in Proposition \ref{exprT} and for $A$ in
Proposition \ref{propA}, show the strong dependence of the
geometry of $C_KM$ on the choice of the timelike vector field $K$.
In particular, the fibres of $\pi |_{C_KM}$ are totally geodesic
(i.e. $T=0$) if and only if $K$ is parallel, Corollary
\ref{totgeo}. This contrasts with the well-known fact which says
that the fibres of the tangent bundle are always totally geodesic.
In our setting, $T$ and $A$ are different from zero. Recall that
semi-Riemannian submersions with both O'Neill tensors non-zero are
less studied in the literature. Thus, the Lorentzian metrics we
obtain are neither products nor Kaluza-Klein, in general.

In Section 4, previous results are specialized for 3-dimensional
time-oriented Lorentzian manifolds. In this case, each null
congruence admits an natural action of $\mathbb{S}^1$ and it
becomes a $\mathbb{S}^1$-principal fibre bundle. A natural
question arises in this setting: {\it When is this action
isometric?} An answer will be found in Theorem \ref{conection}
showing that this holds if and only if the normalized vector field
$1/\sqrt{-g(K,K)}\,K$ is parallel or equivalently, the horizontal
distribution defines a connection on the principal circle bundle
$(C_KM,\pi,M,\mathbb{S}^{1})$.

We end this paper constructing spacetimes from the null congruence
procedure when the base Lorentzian manifold is the 3-dimensional
Minkowski spacetime $\mathbb{L}^3$. According to Remark 2,
$C_K\mathbb{L}^3$ will be stably causal for any timelike vector
field $K$ on $\mathbb{L}^3$. Even more, a suitable choice of such
$K$ will produce nice physically realistic spacetimes of type
$C_K\mathbb{L}^3$. Thus, Theorem \ref{physics} summarizes the main
properties of the obtained examples as follows.

\begin{quote}
For any non-vanishing smooth function $f=f(z)$, $z\in \mathbb{R}$,
the null congruence $C_K\mathbb{L}^3$, with $K=f\partial_z$, is
stably causal, admits an isometric spacelike circle action and is
semi-symmetric (i.e. $R.R=0$). The spacetime $C_K\mathbb{L}^3$ is
of type D in Petrov's classification and it is filled with an
anisotropic perfect fluid. Moreover, $C_K\mathbb{L}^3$ satisfies
the timelike convergence condition if and only if
$2(f')^2-f''f\geq 0$. \end{quote}

It should be remarked that the semi-symmetric notion, $R.R=0$, can
be geometrically interpreted in terms of the invariance, in first
approximation, of the sectional curvature of every tangent plane
after parallel transport along every coordinate parallelogram
\cite{haesen1}. Clearly, it is strictly more weak than the locally
symmetric condition $\nabla R=0$, and so it includes spaces of
constant sectional curvature. Our technique also permits us to
construct a family of semi-symmetric spacetimes $C_K\mathbb{L}^3$,
which are not locally symmetric, for suitable choices of the
vector field $K$, Theorem \ref{semi-symmetric}.

In view of these results and the general approach we show here, it
seems that the construction of spacetimes from the null congruence
is useful to provide new examples of spacetimes which admit a
spacelike circle action.


\section{Definition and properties of the null congruence}

\subsection{Preliminaries}

Let $(M,g)$ be an $n(\geq 3)$-dimensional Lorentzian manifold,
that is, a (connected) smooth manifold $M$ endowed with a
nondegenerate metric $g$ with signature $(-,+,\ldots,+)$. We shall
write $\nabla$ for its Levi-Civita connection, $R$ for its
Riemann-Christoffel curvature tensor\footnote{Our convention on
the curvature tensor is $R(X,Y)Z=
\nabla_{X}\nabla_{Y}Z-\nabla_{Y}\nabla_{X}Z-\nabla_{\left[X,Y\right]}Z.$}
and $\mathrm{Ric}$ for its Ricci tensor. Remember that a  tangent
vector $u\in T_{p}M$ is said to be timelike if $g(u,u)<0$, null if
$g(u,u)=0$ and $u\neq 0$, and spacelike if $g(u,u)>0$ or $u=0$. A
vector field $K\in\mathfrak{X}(M)$ is said to be timelike if
$K_{p}$ is timelike for all $p\in M$. From now on, the Lorentzian
manifold $(M,g)$ is assumed to be time-oriented, that is, a global
timelike vector field $K$ has been fixed and so a timelike or null
tangent vector $v\in T_{p}M$ is said to be future (resp. past)
with respect to K if $g(v,K_{p})<0$ (resp. $g(v,K_{p})>0$).

The null congruence associated with $K$ is defined as the set

\[ C_{K}M = \left\{v\in TM \mid g(v,v)=0\ \mbox{and}\ g(v,K_{\pi(v)})=-1\right\}\ , \]

{\noindent}with $\pi: TM\rightarrow M$ the natural projection. For each $p\in M$ we
put $(C_{K}M)_{p} = C_{K}M\cap T_{p}M$, and thus $C_{K}M =\bigcup_{p\in
M}(C_{K}M)_{p}$. Note that we take the null congruence in the future null cone, in
contrast to the definition used in \cite{gut1,harris,koch}, and that the number $-1$
is only a normalization and any other negative real could be used to define
$C_{K}M$.

We next recall that $C_{K}M$ is an orientable submanifold of $TM$
with dimension $2(n-1)$ and $(C_{K}M,\pi,M,\S^{n-2})$ is a
spherical fibre bundle with structure group $O(n-1)$, proofs of
these facts are given in \cite{gut1}.

Let $(M,g)$ be a time-orientable Lorentzian manifold. If $K$ and
$\mathcal{T}$ are arbitrary timelike vector fields on $M$, then we
have the bundle diffeomorphism $\sigma:C_{K}M\rightarrow
C_{\mathcal{T}}M$ given as follows,
\begin{equation} \label{sigma}\sigma(v)=-\frac{v}{g\big(v,\mathcal{T}_{\pi(v)}\big)}.
\end{equation} Therefore, all null
congruences on $(M,g)$ are diffeomorphic manifolds and they could
be seen as its manifold of null directions.

Recall that the connection map or connector $c$ associated with the Levi-Civita
connection $\nabla$ is defined by
\[ c: TTM\rightarrow TM\, ,\ \ \ \ X\mapsto  \frac{\nabla\alpha}{\mbox{d}t}\mid_{0}\ , \]
{\noindent}where $\alpha$ is a curve in $TM$ with $\alpha^{\prime}(0)= X$ and
$\frac{\nabla\alpha}{\mbox{d}t}$ is the covariant derivative of the vector field
$\alpha$ along the curve $\pi\circ\alpha$ on $M$.

For each $v\in TM$, the canonical identification between the
tangent spaces will be denoted by $$(\,\, )_{v}:
T_{\pi(v)}M\rightarrow T_{v}T_{\pi(v)}M\ ,\ \ \ \ u\mapsto
(u)_{v},$$ and so $c((u)_{v})=u$. If $X\in\mathfrak{X}(M)$, the
vector field $\X\in\mathfrak{X}(TM)$, given by
$\X_{v}=(X_{\pi(v)})_{v}$, is called the vertical lift of $X$ to
$TM$. Recall further that the position vector field, or Liouville
field, $\A\in\mathfrak{X}(TM)$ is given by $\A_{v}= (v)_{v}$.
These vector fields are characterized by the equalities,
$$
\X \omega=\omega(X) \circ \pi,\,\,\,\,\,\,\A\omega=\omega,
$$
for every $\omega\in \mathfrak{X}^{\star}(M)$.

\begin{remark} {\rm
We assume that $M$ is oriented and $\Omega_g$ is the volume form
of the oriented Lorentzian manifold $(M,g)$. Let $\Psi$ be the
$(n-2)$-form on $C_{K}M$ given by,
$$
\Psi(\xi_{1},...,\xi_{n-2})=\Omega_{g}(U_{\pi
(v)},c(\xi_{1}),...,c(\xi_{n-2}),v),
$$
where $\xi_{1},...,\xi_{n-2}\in T_{v}C_{K}M$ and $U=\frac{1}{\sqrt{-g(K,K)}}K$. It
is not difficult to check that its restriction to every fibre $(C_{K}M)_p$ provides
us an orientation and so the spherical fibre bundle $(C_{K}M,\pi,M,\S^{n-2})$ is
oriented in the sense of \cite[Chapter 7]{GreHalVan72}. Hence, when $M$ is assumed
to be compact, the de Rham cohomology groups of $M$ and $C_{K}M$ are related by the
Gysin sequence \cite[Chapter 8]{GreHalVan72} as follows,
$$
...\rightarrow H^{k}_{\textrm{de R}}(M)\rightarrow
H^{k+n-1}_{\textrm{de R}}(M)\rightarrow H^{k+n-1}_{\textrm{de
R}}(C_{K}M)\rightarrow H^{k+1}_{\textrm{de R}}(M)\rightarrow ...
$$ and, as usual, the Euler class $\mathcal{X}_{C_{K}M}\in
H^{n-1}_{\textrm{de R}}(M)$ can be considered. Note that the
vanishing of $\mathcal{X}_{C_{K}M}$ does not imply here the
existence of a cross section (a null tangent vector field on $M$)
for $(C_{K}M,\pi,M,\S^{n-2})$. This fact is a remarkable
difference with the Euler class of the spherical fibre bundle
associated to a compact Riemannian manifold. }
\end{remark}

From the Lorentzian metric $g$ on $M$, the Sasaki metric
$\widehat{g}$ on $TM$ is defined as

\[ \widehat{g}(\zeta,\xi) = g(\zeta_{\star},\xi_{\star}) + g(c(\zeta),c(\xi))\ , \]

{\noindent}for all $\zeta,\xi\in TTM$ with
$\zeta_{\star}=\mbox{d}\pi(\zeta)$ and
$\xi_{\star}=\mbox{d}\pi(\xi)$. Recall that $\pi:
(TM,\widehat{g})\rightarrow (M,g)$ is a semi-Riemannian submersion
in the terminology of Gray \cite{gray} and O'Neill \cite{oneill}.
Given $X\in\mathfrak{X}(M)$, the horizontal lift $\widehat{X}\in
\mathfrak{X}(TM)$ of $X$ along $\pi$ is characterized by,
\begin{equation}
\widehat{X}\omega = \nabla_{X}\omega\ , \label{horlift}
\end{equation}
for each one-form $\omega\in\mathfrak{X}^{\star}(M)$ and satisfies
$\mbox{d}\pi(\widehat{X})=X\circ \pi$ and $c(\widehat{X})=0$. Since $(M,g)$ is a
Lorentzian manifold $(TM,\widehat{g})$ is a semi-Riemannian manifold with index 2.


\subsection{Horizontal and vertical vectors for the null congruence}

In the following, $\widehat{g}$ will also represent the induced
tensor on $C_{K}M$ from the Sasaki one of $TM$, unless otherwise
stated. As it was shown in \cite{gut1}, $(C_{K}M,\widehat{g})$ is
a Lorentzian manifold and
$\pi\mid_{C_{K}M}:(C_{K}M,\widehat{g})\rightarrow (M,g)$ is a
semi-Riemannian submersion with spacelike fibres, that is, for
each $p\in M$ the induced tensor
$\widehat{g}_{p}=\widehat{g}\mid(C_{K}M)_p$ turns
$((C_{K}M)_{p},\widehat{g}_{p})$ into a Riemannian manifold.

As usual, the vectors of $TC_{K}M$ tangent to the fibres are called vertical while
the vectors normal to the fibres are called horizontal. At each $v\in(C_{K}M)_{p}$,
we denote by $\mathcal{V}_v$ the tangent space $T_{v}(C_{K}M)_p$ and $\mathcal{H}_v$
the orthogonal complement to $\mathcal{V}_v$ in $T_{v}C_{K}M$. We will write
$\mathcal{V}$ and $\mathcal{H}$ for the corresponding distributions on $C_{K}M$ and
also the orthogonal projections onto them.

Consider the differentiable map,
$$F : TM\rightarrow \R^{2}\ ,\ \ \ \ v\mapsto \Big(\frac{1}{2}g(v,v),\,
g(v,K_{\pi(v)})\Big).$$ Taking into account that $(0,-1)$ is a regular value of $F$
and $F^{-1}\{(0,-1)\}=C_{K}M$, a standard computation gives us, for $v\in C_{K}M$
with $\pi(v)=p$,
$$
T_{v}C_{K}M=\Big\{\xi\in T_{v}TM: g(c(\xi),v)= g(c(\xi),K_{p})
+g(v,\nabla_{\xi_{\star}}K)=0\Big\},
$$
\begin{equation}
\mathcal{V}_{v}=\Big\{(u)_{v}\in T_{v}T_{p}M:
g(u,v)=g(u,K_{p})=0\Big\}, \label{eq:Vperp1}
\end{equation}
and
\begin{equation}
\mathcal{H}_{v}=\Big\{\xi\in
T_{v}TM:c(\xi)=g(v,\nabla_{\xi_{\star}}K)v\Big\}.
\label{eq:hordistr}
\end{equation}
As a direct consequence, the horizontal lift along the semi-Riemannian submersion
$\pi\mid_{C_{K}M}$ of $X\in \mathfrak{X}(M)$ is the vector field
$\widetilde{X}\in\mathfrak{X}(C_{K}M)$, given for every $v\in C_{K}M$ by,
\begin{equation}
\widetilde{X}_{v} = \widehat{X}_{v} +g\big(v,\nabla_{X_{\pi(v)}}K\big)\A_{v}\ ,
\label{eqcor1}
\end{equation}
{\noindent}where $\widehat{X}$ is the horizontal lift of $X$ to
$TM$ given in \rm{(\ref{horlift})}.

The orthogonal subspace of $\mathcal{V}_{v}$ in $T_{v}T_{p}M$ is,
\begin{equation}
\mathcal{V}_{v}^{\perp}=\mbox{Span}\left\{
\xi_{1}(v),\xi_{2}(v)\right\} \subset T_{v}T_{p}M,
\label{eq:Vperp}
\end{equation}
whereby
\begin{eqnarray}
\xi_{1} = \frac{\K}{\sqrt{-\widehat{g}(\K,\K)}} \quad \mbox{and}
\quad \xi_{2} = \sqrt{-\widehat{g}(\K,\K)}\A - \xi_{1}\ ,
\label{xi1xi2}
\end{eqnarray}
which satisfy $\widehat{g}(\xi_{1},\xi_{1})\, =\, -1\, =\, -
\widehat{g}(\xi_{2},\xi_{2})$ and $ \widehat{g}(\xi_{1},\xi_{2})\, =\, 0$.

For every $X\in\mathfrak{X}(M)$, let $\X^{T}$ be the tangent part
to $C_{K}M$ of $\X$. Since, $\X_{v}\in
T_{v}T_{p}M=\mathcal{V}_{v}\oplus\mathcal{V}_{v}^{\perp}$ for all
$v\in C_{K}M$, it follows from (\ref{eq:Vperp}) that,
\begin{equation}
\X^{T} = \X + \left[ g(K,K)\circ\pi\cdot X^{\flat} +
K^{\flat}(X)\circ\pi\right]\A +X^{\flat}\K\ , \label{eqlemma1}
\end{equation}
where $X^{\flat}$ is the one-form $g$-equivalent to the vector
field $X$, i.e., $X^{\flat}(Y)=g(X,Y)$, for all
$Y\in\mathfrak{X}(M)$.


\begin{remark} {\rm
Two causality conditions which are frequently used in the study of
Lorentzian manifolds are stable causality and the global
hyperbolicity condition. A \textit{stably causal manifold} $(M,g)$
is characterized by the existence of a function $f$ on $M$ whose
gradient is everywhere timelike \cite[Proposition 6.4.9]{hawking}.
A Cauchy hypersurface in $M$ is a subset which is met exactly once
by every inextendible causal curve in $M$. Recently, the
\textit{globally hyperbolic manifolds} are characterized by the
existence of a \textit{smooth} Cauchy hypersurface \cite{bernal}.

Every null congruence on a stably causal Lorentzian manifold
inherits this condition. In fact, taking into account that
$\pi\mid_{C_{K}M}$ is a semi-Riemannian submersion, it follows
that if $f\in C^{\infty}(M)$ has a timelike gradient, then
$f\circ\pi\mid_{C_{K}M}$ has also a timelike gradient. The
question remains open if the global hyperbolicity condition is
preserved by the null congruence construction. }
\end{remark}


\section{The O'Neill fundamental tensors of the null congruence}

In this section we study the relation between the geometries of a
Lorentzian manifold and a null congruence over this manifold. This
relation is expressed through the knowledge of the O'Neill
fundamental $(1,2)$-tensors $T$ and $A$ of the semi-Riemannian
submersion $\pi\mid_{C_{K}M}$. Recall that $T$ and $A$ are given
by,
\begin{eqnarray}
T_{E_{1}}E_{2} & = & \mathcal{H}\widehat{\nabla}_{\mathcal{V}E_{1}}\mathcal{V}E_{2} + \mathcal{V}\widehat{\nabla}_{\mathcal{V}E_{1}}\mathcal{H}E_{2}\ , \label{defT} \\
A_{E_{1}}E_{2} & = &
\mathcal{H}\widehat{\nabla}_{\mathcal{H}E_{1}}\mathcal{V}E_{2} +
\mathcal{V}\widehat{\nabla}_{\mathcal{H}E_{1}}\mathcal{H}E_{2}\ ,
\nonumber
\end{eqnarray}

{\noindent}whereby $E_{1},E_{2}\in\mathfrak{X}(C_{K}M)$ and
$\widehat{\nabla}$ is the Levi-Civita connection of
$\widehat{g}\mid_{C_{K}M}$. Recall that if
$U,V\in\mathfrak{X}(C_{K}M)$ are vertical vector fields, there
holds that $T_{U}V$ is the second fundamental form of the fibres.
Because $T$ is vertical, i.e., $T_{E}=T_{\mathcal{V}E}$, the
fibres of $\pi$ are totally geodesic if and only if $T=0$. On the
other hand, if $X,Y\in\mathfrak{X}(C_{K}M)$ are horizontal, there
holds that,
\begin{equation}
A_{X}Y=\frac{1}{2}\mathcal{V}\left[X,Y\right]. \label{exprA}
\end{equation}
Because $A$ is horizontal, i.e., $A_{E}=A_{\mathcal{H}E}$, the O'Neill fundamental
tensor $A$ characterizes the integrability of the horizontal distribution
$\mathcal{H}$, \cite{oneill}.

For each $z\in T_{p}M$ we can construct the following linear form
on $T_{p}M$, $\Gamma_{z}(u) = g(u,\nabla_{z}K)$, $u\in T_{p}M$.
Then, if $Z\in\mathfrak{X}(M)$, we can consider
$\Gamma_{Z}\in\mathfrak{X}^{\star}(M)$ with
$\Gamma_{Z}(u)=g(u,\nabla_{Z_{\pi(u)}}K)$. In particular, from
(\ref{eqcor1}), it follows that

\begin{equation}
\widetilde{Z}\, =\, \widehat{Z}\, +\, \Gamma_{Z}\, \A\ . \label{horliftZ}
\end{equation}

We are now able to calculate the O'Neill fundamental tensor $T$.

\begin{proposition} \label{exprT}
Let $v\in C_{K}M$ and $\xi,\zeta\in \mathcal{V}_{v}$. The
fundamental tensor $T$ of the semi-Riemannian submersion
$\pi\mid_{C_{K}M}$ satisfies the condition

\begin{equation}
\widehat{g}(T_{\xi}\zeta,\mathcal{Z}) = -\Gamma_{\mathcal{Z}_{\star}}(v)\ \widehat{g}(\xi,\zeta)\ , \label{eqT}
\end{equation}

{\noindent}for every horizontal $\mathcal{Z}\in T_{v}C_{K}M$.
\end{proposition}

{\noindent}\textbf{Proof:} Take $x,y\in T_{p}M$ so that $\xi =
(x)_{v}$, $\zeta =(y)_{v}$ and extend the tangent vectors $x$ and
$y$ to vector fields $X,Y\in\mathfrak{X}(M)$ respectively. By
using (\ref{eq:Vperp1}), we find that $\X^{T}_{v}=\xi$ and
$\Y^{T}_{v}=\zeta$.

From the definition of $T$, we have that,
$$T_{\X^{T}}\Y^{T} =
\mathcal{H}\widehat{\nabla}_{\X^{T}}\Y^{T} .$$ Thus, for every
$Z\in\mathfrak{X}(M)$ we have to calculate
$\widehat{g}(\widehat{\nabla}_{\X^{T}}\Y^{T},\widetilde{Z})(v)$,
where $\widetilde{Z}$ is the horizontal lift of $Z$ along
$\pi\mid_{C_{K}M}$ given in (\ref{horliftZ}).

From the Koszul formula we find,

\begin{eqnarray*}
\lefteqn{2\, \widehat{g}(\widehat{\nabla}_{\X^{T}}\Y^{T},\widetilde{Z}) = \X^{T}\widehat{g}(\Y^{T},\widetilde{Z}) +\Y^{T}\widehat{g}(\X^{T},\widetilde{Z}) -\widetilde{Z}\, \widehat{g}(\X^{T},\Y^{T})} \\
& &
-\widehat{g}\left(\X^{T},\left[\Y^{T},\widetilde{Z}\right]\right)
+\widehat{g}\left(\Y^{T},\left[\widetilde{Z},\X^{T}\right]\right)
+\widehat{g}\left(\widetilde{Z},\left[\X^{T},\Y^{T}\right]\right)\
.
\end{eqnarray*}

{\noindent}The horizontal and vertical vector fields are
$\widehat{g}$-orthogonal and the distribution formed by the
vertical vector fields of every submersion is integrable. Thus, by
using Frobenius' theorem, the Koszul formula simplifies to,

\begin{equation}
2\, \widehat{g}(\widehat{\nabla}_{\X^{T}}\Y^{T},\widetilde{Z})
 = -\widetilde{Z}\, \widehat{g}(\X^{T},\Y^{T}) +\widehat{g}\left(\X^{T},\left[\widetilde{Z},\Y^{T}\right]\right) +\widehat{g}\left(\Y^{T},\left[\widetilde{Z},\X^{T}\right]\right)\ . \label{propTeq1}
\end{equation}

On the other hand, from (\ref{eqlemma1}) we have that

\[ \left[\widetilde{Z},\X^{T}\right] = \left[\widetilde{Z},\X\right] +\Theta_{X}\left[\widetilde{Z},\A\right] +\widetilde{Z}\Theta_{X}\cdot\A +X^{\flat}\left[\widetilde{Z},\K\right] +\widetilde{Z}X^{\flat}\cdot\K\ , \]
where, $ \Theta_{X} = g(K,K)\circ\pi\cdot X^{\flat} +
g(K,X)\circ\pi\ \in\ C^{\infty}(TM)$.

{\noindent}Taking into account that
$\Y^{T}\in\mathfrak{X}(C_{K}M)$ is vertical, we find

\begin{equation}\label{9}
\widehat{g}\left(\Y^{T},\left[\widetilde{Z},\X^{T}\right]\right) =
\widetilde{g}\left(\Y^{T},\left[\widetilde{Z},\X\right]
+\Theta_{X}\left[\widetilde{Z},\A\right]
+X^{\flat}\left[\widetilde{Z},\K\right]\right).
\end{equation}
{\noindent}For every $\omega\in\mathfrak{X}^{\star}(M)$, using
(\ref{horliftZ}), we have

\begin{eqnarray*}
\left[\widetilde{Z},\X\right]\omega & = & Z(\omega(X))\circ\pi -\X(\nabla_{Z}\omega+\Gamma_{Z}\cdot\omega) \\
 & = & \omega(\nabla_{Z}X)\circ\pi -(\Gamma_{Z}(X)\circ\pi)\cdot\omega -\Gamma_{Z}\cdot(\omega(X)\circ\pi)\ ,
\end{eqnarray*}

{\noindent}or

\[ \left[\widetilde{Z},\X\right]= \D(Z,X) -(\Gamma_{Z}(X)\circ\pi)\A -\Gamma_{Z}\, \X\ , \]

{\noindent}where $\D(Z,X)$ is the vertical lift of $\nabla_{Z}X$
to $TM$. Therefore,

\[ \widehat{g}\left(\Y^{T},\left[\widetilde{Z},\X\right]\right) = \widehat{g}\Big(\Y^{T},\D(Z,X) -\Gamma_{Z}\X\Big)\ , \]

{\noindent}and analogously,

\[ \widehat{g}\left(\Y^{T},\left[\widetilde{Z},\K\right]\right) = \widehat{g}\Big(\Y^{T},\D(Z,K)\Big)\ . \]

{\noindent}Further, using the fact that
$\left[\widetilde{Z},\A\right] = -\Gamma_{Z}\, \A$, we find from
(\ref{9}) that,

\[ \widehat{g}\left(\Y^{T},\left[\widetilde{Z},\X^{T}\right]\right) = \widehat{g}\Big(\Y^{T},
\D(Z,X) -\Gamma_{Z}\, \X\Big) + X^{\flat}\,
\widehat{g}\Big(\Y^{T},\D(Z,K)\Big)\ ,
\]

{\noindent}and thus, taking into account that $g(X_{p},v)=0$, we
get,

\begin{equation}\label{10}
\widehat{g}\left(\Y^{T},\left[\widetilde{Z},\X^{T}\right]\right)(v)
= g(Y_{p},\nabla_{Z_{p}}X) -\Gamma_{Z}(v)\, g(X_{p},Y_{p}).
\end{equation}

Since $\Theta_{X}$, $\Theta_{Y}$, $\widehat{g}(\X,\A)$,
$\widehat{g}(\Y,\A)$, $\widehat{g}(\X,\K)$, $\widehat{g}(\Y,\K)$,
$X^\flat$, and $Y^\flat$ vanish at $v$, it is not difficult to
check that,

\begin{equation}\label{11}
\widetilde{Z}\, \widehat{g}(\X^{T},\Y^{T})(v) = Z_{p}\, g(X,Y).
\end{equation}

{\noindent}Finally, from (\ref{9}), (\ref{10}) and (\ref{11}), and
using the fact that the Levi-Civita connection is metric
compatible, the result follows from (\ref{propTeq1}). \hfill
$\Diamond$

\vspace{0.5cm}

As a direct consequence we find the following result.

\begin{corollary} \label{totgeo}
The fibres of the semi-Riemannian submersion $\pi\mid_{C_{K}M}$
are totally geodesic if and only if the timelike vector field $K$
is parallel.
\end{corollary}

We will now calculate the O'Neill fundamental tensor $A$ of
$\pi\mid_{C_{K}M}$.

\begin{proposition} \label{propA}
For every $X,Y\in\mathfrak{X}(M)$, the fundamental tensor $A$ of
the semi-Riemannian submersion $\pi\mid_{C_{K}M}$ satisfies

\[ 2\,A_{\widetilde{X}}\widetilde{Y} = -(R_{XY})^{-} -\widehat{g}((R_{XY})^{-},\K)\A, \]
where $(R_{XY})^{-}\in \mathfrak{X}(TM)$ is given by $(R_{XY})^{-}_{v}=
(R_{X_{p}Y_{p}}v)_{v}$ with $p=\pi(v)$.

\end{proposition}

{\noindent}\textbf{Proof:} From (\ref{exprA}) we know that
$A_{\widetilde{X}}\widetilde{Y} =
\frac{1}{2}\mathcal{V}\left[\widetilde{X},\widetilde{Y}\right]$.
Since $[\widehat{X},\A]=[\widehat{Y},\A]=0$, using
(\ref{horliftZ}), we find,

\[ \left[\widetilde{X},\widetilde{Y}\right] =
\left[\widehat{X},\widehat{Y}\right]
+\Big(\widehat{X}\Gamma_{Y}-\widehat{Y}\Gamma_{X}\Big)\A\ . \]

Let $\mathcal{V}^{\star}$ be the vertical projection of the
semi-Riemannian submersion $\pi:(TM,\widehat{g})\rightarrow
(M,g)$. It is well-known from \cite{gray} that
$\mathcal{V}^{\star}\left[\widehat{X},\widehat{Y}\right] =
-(R_{XY})^{-}$ and therefore,
\[ \mathcal{V}^{\star}\left[\widetilde{X},\widetilde{Y}\right]
= -(R_{XY})^{-} +\Big(\widehat{X}\Gamma_{Y}
-\widehat{Y}\Gamma_{X}\Big)\A.
\]
Since
$\mathcal{V}^{\star}\left[\widetilde{X},\widetilde{Y}\right](v)\in
T_{v}T_{p}M= \mathcal{V}_{v}\oplus\mathcal{V}_{v}^{\perp}$ for all
$v\in C_{K}M$, its vertical part with respect to the null
congruence is given by,

\begin{equation}\label{12}
\mathcal{V}\left[\widetilde{X},\widetilde{Y}\right]=
\mathcal{V}^{\star}\left[\widetilde{X},\widetilde{Y}\right]+
\widehat{g}\Big(\mathcal{V}^{\star}\left[\widetilde{X},\widetilde{Y}\right],\xi_{1}\Big)\xi_{1}-
\widehat{g}\Big(\mathcal{V}^{\star}\left[\widetilde{X},\widetilde{Y}\right],\xi_{2}\Big)\xi_{2}.
\end{equation}
{\noindent}If we take into account the symmetries of the curvature
tensor, we find that $$\widehat{g}((R_{XY})^{-},\A)=0,$$ and so a
straightforward computation from (\ref{12}) gives us the result.
\hfill $\Diamond$

\begin{corollary} \label{flatcase}
The fundamental tensor $A$ of $\pi\mid_{C_{K}M}$ vanishes if and only if $(M,g)$ is flat.
\end{corollary}

{\noindent}\textbf{Proof:} From Proposition \ref{propA} it follows
that if $R=0$, $A$ vanishes on the horizontal vectors, and hence
identically. Conversely, suppose that $A=0$ and fix $v\in C_{K}M$.
From Proposition \ref{propA} it follows that,

\begin{equation}
R_{x y}v = -g\Big(R_{x y}v,K_{\pi(v)}\Big)\, v\ , \label{corAeq1}
\end{equation}

{\noindent}for all $x,y\in T_{\pi(v)}M$. Let $\Pi$ be a degenerate
plane containing $v$, and suppose that $\Pi=\mbox{Span}\{v,z\}$.
From (\ref{corAeq1}) it follows that the null sectional curvature
of the plane $\Pi$ with respect to $v$ vanishes. Thus, all null
sectional curvatures vanish and therefore the sectional curvature
of $(M,g)$ must be a constant $k$ \cite{harris1}. Suppose now that
$x,y\in (C_{K}M)_{\pi(v)}$, with $x,y,v$ linearly independent. We
can then write (\ref{corAeq1}) as follows,
\[ k\left\{g(v,y)x -g(v,x)y\right\} = k \left\{g(v,y) -g(v,x)\right\} v\ . \]
Thus, $k\, g(v,x) = k\, g(v,y) = 0$. But, two orthogonal null
vectors must be collinear, hence $k=0$. \hfill $\Diamond$


\section{The 3-dimensional case}

\subsection{General facts}

Let $(M,g)$ be an oriented $3$-dimensional Lorentzian manifold
which admits a timelike vector field $K\in \mathfrak{X}(M)$. We
normalize $K$ in order to obtain a unitary timelike vector field
$\mathcal{T}=f K$, where $f=\frac{1}{\sqrt{-g(K,K)}}$.

We can introduce an action of the Lie group $\S^{1}$ on
$C_{\mathcal{T}}M$ as follows,
$$
\S^{1}\times C_{\mathcal{T}}M\longrightarrow
C_{\mathcal{T}}M,\,\,\,\,\,(e^{\bf{i}\theta},w)\mapsto
e^{\bf{i}\theta}\cdot
w=\mathcal{T}_{\pi(w)}+\cos\theta\,\bar{w}+\sin\theta\,\bar{u},
$$
where $w=\mathcal{T}_{\pi(w)}+\bar{w}$, with $\bar{w}\in
\mathcal{T}^{\perp}$ and
$\mathcal{B}=\{T_{\pi(w)},\bar{w},\bar{u}\}$ is an oriented
$g$-orthonormal basis for $T_{\pi(w)}M$. It is not difficult to
show that $(C_{\mathcal{T}}M, \pi, M,\S^{1})$ is a principal fibre
bundle.

Now an action of the Lie group $\S^{1}$ on $C_{K}M$ can be
introduced, by using the bundle diffeomorphism $\sigma$ given in
(\ref{sigma}), as follows,
$$
\Phi:\S^{1}\times C_{K}M\longrightarrow
C_{K}M,\,\,\,\,\,\Phi(e^{\bf{i}\theta},v)=\sigma^{-1}\Big(e^{\bf{i}\theta}\cdot\sigma(v)\Big)=f(\pi(v))\Big(e^{\bf{i}\theta}\cdot\sigma(v)\Big).
$$
Hence $(C_{K}M,\pi,M,\S^{1})$ becomes a principal fibre bundle for
every timelike vector field $K$. As usual, we denote
$\Phi(e^{\bf{i}\theta},v)=\Phi_{\theta}(v)$.

\begin{theorem}\label{conection}
For the action $\Phi$ defined above, the following are equivalent:
\begin{enumerate}
\item $\mathcal{T}$ is a parallel vector field, \item The
horizontal distribution $\mathcal{H}$ of the semi-Riemannian
submersion, $$\pi:(C_{K}M,\widehat{g})\rightarrow (M,g),$$ defines
a connection on the principal fibre bundle
$(C_{K}M,\pi,M,\S^{1})$, \item The action $\Phi$ is isometric for
$\widehat{g}\mid_{C_{K}M}$.
\end{enumerate}
\end{theorem}

{\noindent}\textbf{Proof:} Let $v\in (C_{K}M)_{p}$, the horizontal
distribution $\mathcal{H}$ defines a connection if for each
$e^{\bf{i}\theta}\in \S^{1}$,

\begin{equation}
d\Phi_{\theta}(X)\in \mathcal{H}_{\Phi_{\theta}(v)}\ ,
\label{eq:proof1}
\end{equation}

{\noindent}with $X\in \mathcal{H}_{v}$. Let $\lambda$ be a curve
in $C_{K}M$ with $\lambda(0)=v$ and $\lambda'(0)=X$. A direct
computation yields that,

\begin{equation}
c\Big(d\Phi_{\theta}(X)\Big) = (X_{\star}f)\,
(e^{\bf{i}\theta}\cdot\sigma(v)) + f(p)\,
\frac{\nabla(e^{\bf{i}\theta}\cdot(\sigma\circ\lambda))}{dt}\mid_{0}\
. \label{eq:connector}
\end{equation}

{\noindent}From (\ref{eq:connector}) and (\ref{eq:hordistr}) it
follows immediately that (\ref{eq:proof1}) is equivalent to,

\begin{equation}\label{2}
g\Big(e^{\bf{i}\theta}\cdot\sigma(v),
\nabla_{X_{\star}}\mathcal{T}\Big)\,
\big(e^{\bf{i}\theta}\cdot\sigma(v)\big)\ = \ \frac{\nabla
\big(e^{\bf{i}\theta}\cdot(\sigma\circ\lambda)\big)}{dt}\mid_{0}\
.
\end{equation}

Take the vector fields $\bar{V},\bar{U}\in \mathfrak{X}(\pi\circ
\lambda)$ such that $\sigma \circ \lambda=\mathcal{T}+\bar{V}$,
with $\bar{V}\in \mathcal{T}^{\perp}$, and
$\{\mathcal{T},\bar{V},\bar{U}\}$ is an oriented orthonormal basis
along $\pi \circ \lambda$. Then,

$$ e^{\bf{i}\theta}\cdot(\sigma\circ\lambda) = \mathcal{T} +\cos\theta\,
\bar{V} +\sin\theta\, \bar{U}\ . $$

{\noindent}Equation (\ref{2}) is now equivalent to,
$$
\Big(\cos\theta\,
g(\bar{v},\nabla_{X_{\star}}\mathcal{T})+\sin\theta\,
g(\bar{u},\nabla_{X_{\star}}\mathcal{T})\Big)(\mathcal{T}_{p}+\cos\theta\,
\bar{v}+\sin\theta\, \bar{u}) =
$$
\begin{equation}\label{3}
\nabla_{X_{\star}}\Big(\mathcal{T}+\cos\theta\,
\bar{V}+\sin\theta\, \bar{U}\Big),
\end{equation}
where $\bar{V}(0)=\bar{v}$ and $\bar{U}(0)=\bar{u}$. Taking into account that $X$ is
a horizontal tangent vector, we obtain that,
\begin{equation}\label{4}
g(v,\nabla_{X_{\star}}\mathcal{T})v=f(p)^{2}\,\Big(
\nabla_{X_{\star}}\mathcal{T} +\nabla_{X_{\star}}\bar{V}\Big)\ .
\end{equation}
Assume that $\mathcal{T}$ is a parallel vector field, from
(\ref{4}) it is easily deduced that $\nabla_{X_{\star}}\bar{V}=0$
and hence also $\nabla_{X_{\star}}\bar{U}=0$. Using (\ref{3}) we
complete the proof of $(1)\Rightarrow (2)$.

Conversely, if the equality holds in (\ref{3}) for every
$e^{\bf{i}\theta}\in \S^{1}$, by taking the product with
$\bar{v}$, we find,
$$
\cos^{2}\theta\,
g(\nabla_{X_{\star}}\mathcal{T},\bar{v})+\sin\theta \cos\theta\,
g(\nabla_{X_{\star}}\mathcal{T},\bar{u})=
g\Big(\nabla_{X_{\star}}\mathcal{T},\bar{v}\Big)+\sin\theta\,
g\Big(\nabla_{X_{\star}}\bar{U},\bar{v}\Big)\ ,
$$
and from (\ref{4}) we have,

$$
-\sin^{2}\theta\,
g(\bar{v},\nabla_{X_{\star}}\mathcal{T})-\sin\theta(
1-\cos\theta)\, g(\bar{u},\nabla_{X_{\star}}\mathcal{T})=0\ .
$$
Therefore, $g(\bar{v},\nabla_{X_{\star}}\mathcal{T})=
g(\bar{u},\nabla_{X_{\star}}\mathcal{T})=0$ and $\mathcal{T}$ is a
parallel vector field.

Clearly, if $\Phi$ is an isometric action, then
$d\Phi_{\theta}(\mathcal{H}_{v})=\mathcal{H}_{\Phi_{\theta}(v)}$
and so $(3)\Rightarrow(2)$.

Finally, if we assume that $\mathcal{T}$ is a parallel vector
field, the horizontal distribution is a connection and therefore
it can be easily deduced from (\ref{eq:proof1}) that the action
$\Phi$ is isometric on horizontal tangent vectors.

On the other hand, if $\xi\in \mathcal{V}_{v}$ and $\alpha$ is a
curve in $(C_{K}M)_{p}$, with $\alpha(0)=v$ and $\alpha'(0)=\xi$,
then, in a similar way as in (\ref{eq:connector}) we can show
that,
\begin{equation}\label{7}
g\Big(c(d\Phi_{\theta}(\xi)),c(d\Phi_{\theta}(\xi))\Big) =
f(p)^{2}\, g\Big(\frac{\nabla
(e^{\bf{i}\theta}\cdot(\sigma\circ\alpha))}{dt}\mid_{0},\frac{\nabla
(e^{\bf{i}\theta}\cdot(\sigma\circ\alpha))}{dt}\mid_{0}\Big)\ .
\end{equation}
Let us fix an oriented orthonormal basis
$\{\mathcal{T}_{p},e_{1},e_{2}\}$ for $T_{p}M$. We can then
decompose $\sigma \circ \alpha=\mathcal{T}_{p}+\bar{V}^{\alpha}$,
where $\bar{V}^{\alpha}=
\bar{V}^{\alpha}_{1}e_{1}+\bar{V}^{\alpha}_{1}e_{2}\in
\mathcal{T}^{\perp}_{p}$. We can further construct a vector
$\bar{U}^{\alpha}=-\bar{V}^{\alpha}_{2}e_{1}+\bar{V}^{\alpha}_{1}e_{2}$
such that $\{\mathcal{T}_{p},\bar{V}^{\alpha},\bar{U}^{\alpha}\}$
is an orthonormal basis of $T_{p}M$. Then,

\begin{equation}
c(\xi)= f(p)\Big(\frac{d\bar{V}^{\alpha}_{1}}{dt}\mid_{0}e_{1}
+\frac{d\bar{V}^{\alpha}_{2}}{dt}\mid_{0}e_{2}\Big)\ .\label{8}
\end{equation}

{\noindent}Using (\ref{7}) and (\ref{8}) one can show that,

$$
g\Big(c(d\Phi^{K}_{\theta}(\xi)),c(d\Phi^{K}_{\theta}(\xi))\Big)\
=\ g(c(\xi),c(\xi))\ ,
$$

{\noindent}which completes the proof of $(1)\Rightarrow(3)$.

\hfill $\Diamond$


\subsection{Null congruences on $\L^{3}$}

In order to obtain a family of examples, we consider the case when
the base manifold $M$ is the 3-dimensional Minkowski space
$\L^{3}$, where the natural coordinate vector field $\partial_{z}$
is unit and timelike. From Remark 2 we find that all the null
congruences $C_{K}\L^{3}$ are 4-dimensional, stably causal,
Lorentzian manifolds and from Corollary \ref{flatcase} it follows
that the fundamental tensor of O'Neill $A$ of every null
congruence on $\L^{3}$ vanishes identically. Therefore, from
\cite[Theorem 9.28]{besse} it follows that the only non-vanishing
components of the curvature tensor $\widehat{R}$ of $C_{K}\L^{3}$
have the form,
\begin{equation}
\widehat{g}(\widehat{R}(X,U)Y,U) = \widehat{g}(T_{U}X,T_{U}Y) -
\widehat{g}((\widehat{\nabla}_{X}T)_{U}U,Y)\ , \label{expr1curv}
\end{equation}
where $X,Y$ are horizontal vector fields and $U$ is the spacelike unitary vector
field which spans the vertical distribution. Remark that our choice of sign of the
curvature tensor is opposite as in \cite{besse}.

Let $\widetilde{X}$ be the horizontal lift of $\partial_{x}$ to
$C_{K}\L^{3}$, and analogously for $\widetilde{Y}$ and
$\widetilde{Z}$. It can be deduced from (\ref{expr1curv}) that the
Ricci tensor $\widehat{\mathrm{Ric}}$ of every null congruence
$C_{K}\L^{3}$ satisfies the following properties,
\begin{equation}\label{13}
\widehat{\mathrm{Ric}}(U,U)=\widehat{\mathrm{Ric}}(\widetilde{X},\widetilde{X})+\widehat{\mathrm{Ric}}(\widetilde{Y},\widetilde{Y})
-\widehat{\mathrm{Ric}}(\widetilde{Z},\widetilde{Z})
\end{equation}
$$
\widehat{\mathrm{Ric}}(\widetilde{X},U)=\widehat{\mathrm{Ric}}(\widetilde{Y},U)=\widehat{\mathrm{Ric}}(\widetilde{Z},U)=0.
$$

\begin{remark} {\rm
There exists no non-flat Einstein null congruence $C_{K}\L^{3}$.
In fact, from (\ref{13}), every Einstein null congruence
$C_{K}\L^{3}$ is Ricci flat and it can be deduced that the
sectional curvature $\widehat{\mathcal{K}}$ of non-degenerate
planes in a Ricci flat null congruence $C_{K}\L^{3}$ satisfies
$\widehat{\mathcal{K}}= 0$. }
\end{remark}

\begin{lemma} \label{lemma:curv}
The only non-vanishing components of the Riemann-Christoffel
curvature tensor of the null congruence $C_{K}\L^{3}$, at a point
$v\in C_{K}\L^{3}$ with $\pi(v)=p$, are given as follows,
\begin{eqnarray}
\lefteqn{ \widehat{g}(\widehat{R}(X,U)Y,U)_{v} = g\left(v,\nabla_{X_{\star}\mid_{p}} \nabla_{Y_{\star}\mid_{p}}K\right) } \nonumber \\
 & & - g\left(v,\nabla_{\nabla_{X_{\star}}Y_{\star}\mid_{p}}K\right)
 + 2 g\left(v,\nabla_{X_{\star}\mid_{p}}K\right) g\left(v,\nabla_{Y_{\star}\mid_{p}}K\right)\ , \label{example1}
\end{eqnarray}
where $X,Y$ are horizontal vector fields and $U$ is the spacelike unitary vector
field which spans the vertical distribution.
\end{lemma}

{\noindent}\textbf{Proof:} From \cite[Theorem 9.18]{besse} it is
well known that $T$ satisfies
$\widehat{g}(T_{U}X,T_{U}Y)=-\widehat{g}(T_{U}T_{U}X,Y)$.
Therefore, Proposition~\ref{exprT} implies that,
\begin{equation}
\widehat{g}\left(T_{U}X,T_{U}Y\right)_{v} =
g\left(v,\nabla_{X_{\star}\mid_{p}}K\right)
g\left(v,\nabla_{Y_{\star}\mid_{p}}K\right)\ . \label{expr1}
\end{equation}
On the other hand, there holds that,
\[ (\widehat{\nabla}_{X}T)_{U}U =
 \widehat{\nabla}_{X}(T_{U}U) - T_{\widehat{\nabla}_{X}U}U - T_{U}(\widehat{\nabla}_{X}U)\ ,\]
and from \cite[Lemma 3]{oneill} and $A=0$, it follows that $\widehat{\nabla}_{X}U =
\mathcal{V}\widehat{\nabla}_{X}U$. Taking into account that $X\, \widehat{g}(U,U) =
2\, \widehat{g}(\widehat{\nabla}_{X}U,U) = 0$, one can deduced that
$\widehat{\nabla}_{X}U=0$. Thus,
\[ \widehat{g}((\widehat{\nabla}_{X}T)_{U}U,Y) =  X\Big(\widehat{g}(T_{U}U,Y)\Big) - \widehat{g}(T_{U}U,\widehat{\nabla}_{X}Y)\ . \]

Assume that $X,Y\in\mathfrak{X}(\L^{3})$ are basic horizontal vector fields, i.e.,
they are $\pi$-related to the vector fields $X_{\star},Y_{\star}\in\mathfrak{X}(M)$.
There holds that $\widehat{\nabla}_{X}Y = \mathcal{H}\widehat{\nabla}_{X}Y$ and thus
$\mbox{d}\pi \Big(\widehat{\nabla}_{X}Y\Big) =
\Big(\nabla_{X_{\star}}Y_{\star}\Big)\circ\pi$, \cite[Lemma 1]{oneill}. From
Proposition \ref{exprT} we then find that,
$$
\widehat{g}(T_{U}U,\widehat{\nabla}_{X}Y)_{v}  =
-g(v,\nabla_{\nabla_{X_{\star}}Y_{\star}\mid_{p}}K)\ . \label{expr2}
$$
Finally, from (\ref{eqcor1}) and (\ref{eqT}), we get,
$$
X\Big(
\widehat{g}(T_{U}U,Y)\Big)=-\nabla_{X_{\star}}\Gamma_{Y_{\star}}-\Gamma_{X_{\star}}\Gamma_{Y_{\star}},
$$
which completes the proof.

\hfill $\Diamond$

\vspace{4mm}

Recall that a semi-Riemannian manifold $(M,g)$ is called locally
symmetric if its Riemann-Christoffel curvature tensor satisfies
the condition $\nabla R=0$. Manifolds which satisfy this
condition, also satisfy the integrability condition $R\cdot R=0$,
i.e.,

\begin{equation}\label{RR}
(R\cdot R)(X_{1},X_{2},X_{3},X_{4};X,Y) =
\end{equation}
$$
-g\Big(R(R(X,Y)X_{1},X_{2})X_{3},X_{4}\Big) -
g\Big(R(X_{1},R(X,Y)X_{2})X_{3},X_{4}\Big)
$$
$$
 -g\Big(R(X_{1},X_{2})R(X,Y)X_{3},X_{4}\Big)-
g\Big(R(X_{1},X_{2})X_{3},R(X,Y)X_{4}\Big) = 0\ ,
$$

{\noindent}with $X_{1},X_{2},X_{3},X_{4},X,Y\in\mathfrak{X}(M)$.
However, the converse is not true in general. A manifold which
satisfies the condition $R\cdot R=0$ is called semi-symmetric. A
geometric interpretation of this class of manifolds is given in
\cite{haesen1}.

\begin{theorem}\label{semi-symmetric}
Let $C_{K}\L^{3}$ be the null congruence associated with a
timelike vector field $K\in\mathfrak{X}(\L^{3})$ which satisfies
the condition $\nabla K = \alpha\, K^{\flat}\otimes K$, with
$\alpha\in C^{\infty}(\L^{3})$. Then, $C_{K}\L^{3}$ is a
semi-symmetric Lorentzian manifold. Moreover, if $K(K(\alpha))\neq
0$, the null congruence $C_{K}\L^{3}$ is not locally symmetric.
\end{theorem}

{\noindent}\textbf{Proof:} From Lemma \ref{lemma:curv} we find,
\begin{equation}
\widehat{g}\Big(\widehat{R}(X,U)Y,U\Big)_{v} =
g(X_{\star}\mid_{p},\nabla\alpha)\, g(Y_{\star}\mid_{p},K)\ ,
\label{eq:proofbonita}
\end{equation}
where $v\in C_{K}\L^{3}$ with $\pi(v)=p$. Let $E_{3}\in
\mathfrak{X}(C_{K}\L^{3})$ the horizontal lift of
$\frac{K}{\sqrt{-g(K,K)}}$ and $\{e_{1},e_{2},e_{3}\}$ an
orthonormal basis of $T_{p}\L^{3}$ adapted to our choice of $K$,
i.e., take as timelike direction $e_{3} = E_{3}(v)$. Let
$\{E_{1},E_{2},E_{3}(v)\}$ be the horizontal lift of
$\{e_{1},e_{2},e_{3}\}$ to $T_{v}C_{K}\L^{3}$. With respect to the
$\widehat{g}$-orthonormal basis $\{E_{1},E_{2},E_{3}(v),U_{v}\}$,
the only component of the curvature tensor $\widehat{R}$ which can
be non-vanishing is,
\[ \widehat{g}(\widehat{R}(E_{3},U)E_{3},U)_{v} = - K_{p}(\alpha) . \]
We can introduce a Newman-Penrose null basis as follows,
\begin{eqnarray*}
\widehat{L} = \frac{1}{\sqrt{2}}(U_{v}+E_{3}(v))\ , & \widehat{N}
= \frac{1}{\sqrt{2}}(U_{v}-E_{3}(v))\ , & \widehat{M} =
\frac{1}{\sqrt{2}}(E_{1}+{\bf i}\, E_{2})\ .
\end{eqnarray*}
The scalars which determine the curvature in the Newman-Penrose formalism are
\cite{stephani},

\begin{eqnarray}
\widehat{\Psi}_{0} = \widehat{\Psi}_{1} = \widehat{\Psi}_{3} =
\widehat{\Psi}_{4} = 0 & \mbox{and} & \widehat{\Psi}_{2} =
\overline{\widehat{\Psi}}_{2} = -2\Lambda = -\frac{1}{6}K(\alpha)\
, \label{eq:petrov} \end{eqnarray}

\begin{eqnarray}
\widehat{\Phi}_{0i} = \widehat{\Phi}_{2i} = 0\ ,\ i=0,1,2 &
\mbox{and} & \widehat{\Phi}_{11} = -\frac{1}{4}K(\alpha)\ .
\label{eq:fluid}
\end{eqnarray}

{\noindent}From the classification given in \cite{haesen} it
follows that $(C_{K}\L^{3},\widehat{g})$ is a semi-symmetric
manifold.

In order to show that the null congruence $C_{K}\L^{3}$ is not
locally symmetric, we point out that from
$\widehat{\nabla}_{E_{3}}U = \mathcal{V}\widehat{\nabla}_{E_{3}}U$
and $E_{3}\, \widehat{g}(U,U)=0$, we deduce that
$\widehat{\nabla}_{E_{3}}U=0$. Further, since
$\widehat{\nabla}_{E_{3}}E_{3} =
\mathcal{H}\widehat{\nabla}_{E_{3}}E_{3}$ and
$E_{3}\widehat{g}(E_{3},E_{3})=0$, we have that
$(\widehat{\nabla}_{E_{3}}E_{3})_{v} \in \mbox{Span}\{
E_{1},E_{2}\}$. Therefore, from (\ref{expr1curv}), a
straightforward computation gives us,
\[ \widehat{g}\Big(
(\widehat{\nabla}_{E_{3}}\widehat{R})(E_{3},U)E_{3},U\Big)_{v}=
E_{3}(v)\Big(\widehat{g}(\widehat{R}(E_{3},U)E_{3},U)\Big) =
-\frac{1}{\sqrt{-g(K_{p},K_{p})}}K_{p}(K(\alpha))\ , \] which
completes the proof.

\hfill $\Diamond$

\vspace{0.5cm}

Recall that a spacetime satisfies the timelike convergence condition if and only if
$\mbox{Ric}(W,W)\geq 0$, for every timelike vector field $W$. This is the
mathematical translation of the fact that on average gravity attracts.

\begin{theorem}\label{physics}
Let $f=f(z)\in C^{\infty}(\L^{3})$ be a non-vanishing function.
Then, the $4$-dimensional, semi-symmetric, Lorentzian manifold
$C_{K}\L^{3}$, with $K=f\,
\partial_{z}$, is stably causal and admits an
isometric spacelike circle action. This spacetime is further of
type D in Petrov's algebraic classification of the Weyl tensor and
is filled with an anisotropic perfect fluid. The null congruence
$C_{K}\L^{3}$ satisfies the timelike convergence condition if and
only if $2(f')^{2}- f''f\geq 0$.
\end{theorem}

{\noindent}\textbf{Proof:} Note that $K$ satisfies the condition
$\nabla K= -f'f^{-2}\, K^{\flat}\otimes K$. From (\ref{eq:petrov})
and (\ref{eq:fluid}) it follows immediately that $C_{K}\L^{3}$ is
an anisotropic perfect fluid of Petrov type D. Recall that
$\widetilde{X}$ denotes the horizontal lift of $\partial_{x}$ to
$C_{K}\L^{3}$, and analogously for $\widetilde{Y}$ and
$\widetilde{Z}$. An unitary timelike vector field
$W\in\mathfrak{X}(C_{K}\L^{3})$ can be written as follows,
\[ W = \alpha\, \widetilde{X}+\beta\, \widetilde{Y}+\gamma\,
\widetilde{Z}+\delta\, U\ . \] From (\ref{eq:proofbonita}) it
deduces that $\widehat{\mbox{Ric}}(\widetilde{X},.) =
\widehat{\mbox{Ric}}(\widetilde{Y},.)=0$ and therefore from
(\ref{13}),
\[ \widehat{\mbox{Ric}}(W,W) = (\gamma^{2}-\delta^{2})\,
\widehat{\mbox{Ric}}(\widetilde{Z},\widetilde{Z}) =
(\gamma^{2}-\delta^{2})\Big\{  2 \left(\frac{f'}{f}\right)^{2}-
\frac{f''}{f}\Big\}\ , \] which completes the proof.

\hfill $\Diamond$

\begin{remark} {\rm
If $f(z)=e^{a z}$, with $a\in\R-\{0\}$, the null congruence
$C_{K}\L^{3}$, with $K=f\partial_{z}$, satisfies the timelike
convergence condition and is not locally symmetric. }
\end{remark}


\bibliographystyle{unsrt}

\vspace{20mm}

Stefan Haesen

Department of Mathematics

Katholieke Universiteit Leuven

Celestijnenlaan 200B

3001 Heverlee, Belgium

\texttt{Stefan.Haesen@wis.kuleuven.be}

\vspace{4mm}

Francisco J. Palomo

Department of Applied Mathematics

Universidad de M\'{a}laga

29071 M\'{a}laga, Spain

\texttt{fjpalomo@ctima.uma.es}

\vspace{4mm}

Alfonso Romero

Department of Geometry and Topology

Universidad de Granada

18071 Granada, Spain

\texttt{aromero@ugr.es}

\end{document}